\documentclass[11pt]{revtex4}
\usepackage{natbib} 
 \usepackage{relsize}
\usepackage{hyperref}
\usepackage{amsmath,amsfonts,amssymb}
\hypersetup{dvips,dvipdfm,colorlinks=true,urlcolor=magenta,filecolor=magenta,linktoc=page,citecolor=red,linkcolor=blue,bookmarks=true}
\usepackage{graphicx,epstopdf}
 
\begin{document}
	\title{Cosmic evolution in $f(T)$ gravity theory}
	
	\author{Akash Bose$^1$\footnote {bose.akash13@gmail.com}}
	\author{Subenoy Chakraborty$^1$\footnote {schakraborty.math@gmail.com}}
	\affiliation{$^1$Department of Mathematics, Jadavpur University, Kolkata-700032, West Bengal, India}
	
	\begin{abstract}
		
		The paper deals with cosmology in modified $f(T)$ gravity theory. With some phenomenological choices for the function $f(T)$ it is possible to have cosmological solutions describing different phases of the evolution of the Universe for the homogeneous and isotropic FLRW model. By proper choice of the parameters involved in the function $f(T)$ and also in the cosmological solutions it is shown that a continuous cosmic evolution starting from the emergent scenario to the present late time acceleration is possible. Finally thermodynamical analysis of $f(T)$ gravity is presented.
		
	\end{abstract}
\maketitle

\maketitle

\section{Introduction}

Since 1998, series of cosmological observations namely TypeIa Supernova Hubble diagram (SNeIa)\cite{Riess:1998cb}-\cite{Perlmutter:1998np}, Cosmic microwave background anisotropies (CMBR)\cite{Spergel:2003cb}, Large Scale Structure (LSS)\cite{Tegmark:2003ud} and Baryon Acoustic Oscillation (BAO)\cite{Eisenstein:2005su} predict that our Universe has been undergoing through an accelerated expansion phase. The standard cosmology in the framework of General Relativity (GR) cannot match the overwhelming abundance of the observational evidences of cosmic speedup. One can resolve this paradox by introducing some exotic matter of large negative pressure (known as dark energy (DE)) but difficult questions arise on its nature and introduce further difficult problems. So naturally, cosmologists speculate that these observational evidences may indicate the first signal for the limitation of the Einstein gravity over cosmological scale. As a consequence, there are several modified gravity theories in the literature of which $f(R)$ gravity theory is widely used\cite{Capozziello:2003tk}-\cite{Carroll:2003wy} and gets much attention.

In $f(R)$ gravity theory, the scalar curvature $R$ in the Einstein-Hilbert action is changed to a suitable function $f(R)$. In another modified gravity theory (which is very popular in recent years) known as teleparallel gravity, instead of curvature, torsion represents the gravitational interaction\cite{Einstein:1928}-\cite{Hayashi:1979qx}. In fact Einstein first proposed this model to unify electromagnetism and gravity over Weitzenböck non-Riemannian manifold. As a result the Levi-Civita connection is replaced by Weitzenböck connection in the underlying Riemann-Cartan space-time. Hence pure geometric nature of gravitational interaction is no longer there rather torsion acts as a force and one may interpret gravity as a gauge theory of the translation group\cite{Arcos:2005ec}. It is to be noted that although there are conceptual directions between these two (i.e. GR and teleparallel) gravity theories, still they have equivalent dynamics at least all the classical level.

Further, a generalization to teleparallel gravity (similar to $f(R)$) has been developed \cite{Ferraro:2006jd}-\cite{Harko:2014sja}by replacing torsion scalar $T$ by a generic function $f(T)$. This modified gravity theory is termed as $f(T)$ gravity theory and Linder coined the name. Consequently, there is no equivalence (with GR) at the classical level i.e. distinct dynamics comes into picture. Moreover, there are two important differences between these two theories namely (i) The field equations in $f(T)$ gravity theory remain second order while one has fourth order equations in $f(R)$\cite{Bengochea:2008gz}, (ii) $f(T)$ gravity theory does not satisfy local Lorentz invariance (which is satisfied by $f(R)$ gravity) so that all 16 components of the vierbien are independent and hence it is not possible to fix six of them by a gauge choice\cite{Li:2010cg}. The dynamical object in this model are the four linearly independent vierbein (tetrad) fields which form the orthogonal bases for the tangent space at each point of space-time. The vierbeins are parallel vector fields, which give the theory the descriptor “teleparallel”. The advantage is that the torsion tensor is formed solely from products of first derivatives of the tetrad. This modified teleparallel gravity provides an alternative to inflation \cite{Ferraro:2006jd}. For details see the review \cite{Cai:2015emx}-\cite{Biswas:2015cva}.

In the literature, there are infinite numbers of cosmological solutions describing some of the phases of evolution of the Universe. However, in recent years there are few works \cite{Chakraborty:2014oya}-\cite{Amoros:2013nxa} in the literature which describe the whole evolution of the Universe both in Einstein gravity and in some modified gravity theories. In this context the motivation of the present work is 3-fold: (i) to examine whether $f(T)$ gravity theory describes a non-singular model of the Universe, (ii) to show its correspondence with Einstein gravity and (iii) to describe a continuous cosmic evaluation starting from early inflationary era to present accelerating phase with an appropriate continuous choice of $f(T)$. The paper is organized as follows: section II deals with basic equations of $f(T)$ gravity, non-singular (emergent) model of Universe in $f(T)$ gravity has been studied in section III. In section IV correspondence between Einstein gravity and $f(T)$ gravity is presented. Section V contains continuous cosmic evolution in $f(T)$ gravity. The paper ends with a summary in section VI.
\section{Basic equations of $f(T)$ gravity}
The action for $f(T)$ gravity can be written as ($\kappa=8\pi G=1$) \cite{Anagnostopoulos:2019miu}-\cite{Chen:2010va}
\begin{equation}\label{eq1}
\mathcal{A}=\frac{1}{2}\int d^4x~|e|~\left[T+f(T)+L_m\right]
\end{equation}

where $T$ is torsion scalar, $f(T)$ is a differentiable function of torsion, $|e|=\det \left(e^A_\mu\right)=\sqrt{-g}$, and $L_m$ is the matter Lagrangian.

The torsion scalar is defined as
\begin{equation}\label{eq2}
T=S_{\sigma}^{~\mu \nu}T_{~\mu \nu}^\sigma
\end{equation}

$T_{~\mu \nu}^\sigma$, the torsion tensor is defined as
\begin{equation}\label{eq3}
T_{~\mu \nu}^\sigma=\Gamma^\sigma_{\nu \mu}-\Gamma^\sigma_{\mu \nu}=e^\sigma_ A\left(\partial_\mu e^A_\nu-\partial_\nu e^A_\mu\right)
\end{equation}

where the Weitzenböck connection $\Gamma^\sigma_{\mu \nu}$ is defined as $\Gamma^\sigma_{\mu \nu}=e^\sigma_ A\partial_\nu e^A_\mu$ and $S^{~\mu \nu}_\sigma$, the super-potential is defined as
\begin{equation}\label{eq4}
S^{~\mu \nu}_\sigma=\frac{1}{2} \left(K^{\mu \nu}_{~~\sigma}+\delta^ \mu_\sigma~ T^{\alpha \nu}_{~~\alpha} -\delta^ \nu_\sigma~ T^{\alpha \mu}_{~~\alpha}\right)
\end{equation}

In teleparallel gravity, orthogonal tetrad components $e_A(x^\mu)$ are considered as dynamical variables and geometrically they form an orthonormal basis for the tangent space at each point $x^\mu$ of the manifold i.e. 
\begin{equation}\label{eq4a}
e_Ae_B=\eta_{_{AB}}=diag(+1,-1,-1,-1)
\end{equation}
Further, in a co-ordinate basis one may write $e_A=e^\mu_A\partial_\mu$ where $e^\mu_A$ are the components of $e_A$, with $\mu=0,1,2,3$ and $A=0,1,2,3$. Note that capital letters refer to the tangent space while Greek indices label coordinates on the manifold. Hence the metric tensor is obtained from the dual vierbein as
\begin{equation}\label{eq5}
g_{\mu \nu}(x)=\eta_{_{AB}}e^A_\mu(x) e^B_\nu(x)
\end{equation}

and $K^{\mu \nu}_{~~\sigma}$, the contortion tensor is given by
\begin{equation}\label{eq6}
K^{\mu \nu}_{~~\sigma}= -\frac{1}{2} \left(T^{\mu\nu} _{~~\sigma}-T^{\nu\mu}_{~~\sigma}-T^{~\mu\nu}_{\sigma}\right)
\end{equation}

In the present work we consider flat, homogeneous and isotropic FLRW (Friedmann-Lemaitre-Robertson-Walker) space-time having line element
%\vspace{-0.2 cm}
\begin{equation}\label{eq7}
ds^2=dt^2-a^2(t)\delta_{ij} dx^i dx^j
\end{equation}
 
where $a(t)$ is scale factor.

 Using (\ref{eq2}), (\ref{eq3}), (\ref{eq4}), (\ref{eq4a}) and (\ref{eq6}) one has
$$T=-6H^2$$

where $H=\frac{\dot{a}}{a}$ is the Hubble parameter and overdot represents differentiation with respect to cosmic time $t$. One may note that during cosmic evolution $T$ is negative.

Varying the above action (\ref{eq1}) the modified Einstein field equations become
\begin{equation}\label{eq8}
e^{-1} \partial_\mu(ee^\rho_A S_\rho^{~\mu \nu})[1+f_T] + e^\rho_A S_\rho^{~\mu \nu} \partial_\mu(T)f_{TT}-[1+f_T]e^\lambda_A T^\rho_{~\mu \lambda} S_\rho^{~\nu \mu}+\frac{1}{4}e^\nu_A[T+f(T)]=4\pi G e^\rho_A  T_\rho^{~\nu}
\end{equation}

where $f_T=\frac{d f}{d T}$, $f_{TT}=\frac{d^2 f}{d T^2}$ and $T_\rho^{~\nu}$ denotes the total matter (i.e. baryonic and dark energy) energy momentum tensor.
	
Then for FLRW model the modified Friedmann equations become
\begin{eqnarray}
H^2&=&\frac{1}{2 f_T+1}\left[\frac{\rho}{3}-\frac{f}{6}\right],\label{eq9}\\
2\dot{H}&=&-\frac{(p+\rho)}{1+f_T+2T f_{TT}}.\label{eq10}
\end{eqnarray}	

where $p$ and $\rho$ are respectively the thermodynamic pressure and density of the matter fluid having the conservation equation
\begin{equation}\label{eq11}
\dot{\rho}+3H(p+\rho)=0
\end{equation}
 
It is to be noted that for various choices of $f(T)$, we have different corresponding gravity theories and it is presented in the following table.
\begin{table}[h]
	\begin{tabular}{|c|c|}
		\hline
		Choice of $f(T)$ & Gravity theory\\\hline
		0& Einstein gravity\\\hline
		Nonzero Constant&Einstein gravity with Cosmological Constant\\\hline
		$f(T)=\alpha T$, $\alpha\neq 0$, a constant&Einstein gravity with reconstruction of gravitational constant\\\hline
		General $f(T)$&Modified gravity theory\\\hline
	\end{tabular}
\end{table}

% Equations (\ref{eq9}) and (\ref{eq10}) can also be written as
%\begin{eqnarray}
%3H^2&=&\frac{\rho-\frac{f}{2}}{1-\frac{1}{6H} f_H},\label{eq12}\\
%2\dot{H}&=&-\frac{\gamma \rho}{1-\frac{1}{12}f_{HH}}\label{eq13}.
%\end{eqnarray}

Assuming fluid to be barotropic in nature with equation of state $p=\omega\rho=(\gamma-1)\rho$, $0\leq\gamma\leq2$  and eliminating $\rho$ from equations (\ref{eq10}) and (\ref{eq11}), one has the differential equation for cosmic evolution as,
\begin{equation}\label{eq14}
2\dot{H}\left(1+f_T+2Tf_{TT}\right)=\frac{\gamma}{2}\left[T(2f_T+1)-f\right].
\end{equation}
\section{A non-singular model in $ f(T)$ cosmology}
In the context of cosmic evolution the most common idea of the beginning of the Universe is the standard big bang model which evolves from a singularity. To overcome this singularity problem the cosmologists have an idea of emergent scenario of the Universe which is a non-singular model and the cosmological solution represents Einstein static model in infinite past.

In cosmic evolution to have an emergent scenario, the evolution of Hubble parameter have the well known differential equation [\citealp{Chervon:2014tra},\citealp{Chakraborty:2014ora},\citealp{Das:2018bxc}]
\begin{equation}\label{eq14a}
\dot{H}=\alpha H+\lambda H^2
\end{equation}

with parameters $\alpha$ and $\lambda$ are restricted as $\lambda<0$ and $\alpha>|\lambda H_0|$ for emergent scenario. Comparing equation (\ref{eq14a}) with the evolution equation (\ref{eq14}) in $f(T)$ cosmology one may choose  $f(T)$ as 
\begin{equation}\label{eq15}
f(T)=b+c\sqrt{|T|}-T+d\sqrt{|T|} \ln \left(|T|\right)
\end{equation}

where $b,~c$ and $d$ are arbitrary constants and hence
\begin{equation}
\dot{H}=\beta H-3\gamma H^2\label{eq16}
\end{equation}

where $\beta=\frac{\gamma b\sqrt{6}}{4d}=\gamma\mu,~\mu=\frac{b\sqrt{6}}{4d}.$ Thus this choice of $f(T)$ in equation (\ref{eq15}) is purely phenomenological with the motivation of obtaining a solution for emergent scenario.

Depending on the sign of $\beta$ (i.e. $\mu$), one gets various solutions as follows:

If $\beta>0$ (i.e. $\mu>0$), then the cosmological solutions are given by              
\begin{eqnarray}\label{eq17}
\frac{H}{H_0}&=&\frac{\beta}{3\gamma H_0-\left(3\gamma H_0-\beta\right)e^{-\beta(t-t_0)}} ,\nonumber\\
\frac{a}{a_0}&=&\left[1+\frac{3\gamma H_0}{\beta} \left(e^{\beta(t-t_0)}-1\right)\right]^{\frac{1}{3\gamma}}.
\end{eqnarray}

%and if $\beta>3\gamma H_0$ (i.e. $b>dH_0$), then the solution has the form
%\begin{eqnarray}\label{eq18}
%\frac{H}{H_0}&=&\frac{\beta}{3\gamma H_0+\left(\beta-3\gamma H_0\right)e^{-\beta(t-t_0)}} ,\nonumber\\
%\frac{a}{a_0}&=&\left[1+\frac{3\gamma H_0}{\beta} \left(e^{\beta(t-t_0)}-1\right)\right]^{\frac{1}{3\gamma}}.
%\end{eqnarray}

One may note that equation (\ref{eq17}) represents big bang solution for $\beta<3\gamma H_0$ (i.e. $\mu<3H_0$) while equation (\ref{eq17}) represents emergent solution for $\beta>3\gamma H_0$ (i.e. $\mu>3H_0$) and $\beta=3\gamma H_0$ (i.e. $\mu=3H_0)$ gives the inflationary solution (i.e. the exponential expansion). In particular if one consider $$\beta=6\gamma H_0~~\mbox{i.e.}~~\mu=6H_0$$

and the cosmological solutions are in the following form
\begin{eqnarray}\label{eq19}
\frac{H}{H_0}&=&2-\left(\frac{a}{a_0}\right)^{-3\gamma}\nonumber\\\mbox{i.e.~}
\frac{H}{H_0}&=&\frac{2}{1+e^{-6\gamma H_0 (t-t_0)}},\nonumber\\
\frac{a}{a_0}&=&\left[\frac{1+e^{6\gamma H_0 (t-t_0)}}{2}\right]^{\frac{1}{3\gamma}}.\nonumber\\\mbox{and~~}
\frac{\rho}{\rho_0}&=&\left(\frac{a}{a_0}\right)^{-3\gamma}.
\end{eqnarray}

For emergent solution represented by equation (\ref{eq19}) one has the following asymptotic behavior

$(i)~a\rightarrow a_0  2 ^{-\frac{1}{3\gamma}}, H\rightarrow0 ~\mbox{as}~t\rightarrow -\infty,$

$(ii)~a\sim a_0  2 ^{-\frac{1}{3\gamma}} , H\sim0~\mbox{as}~t\ll t_0,$

$(iii)~a\sim a_0 2 ^ {-\frac{1}{3\gamma}}e^{2H_0 (t-t_0)} , H\sim2H_0~\mbox{as}~t\gg t_0. $

\begin{figure}[h]
	\begin{minipage}{0.495\textwidth}
		\includegraphics[height=.7\textwidth,width=.95\textwidth]{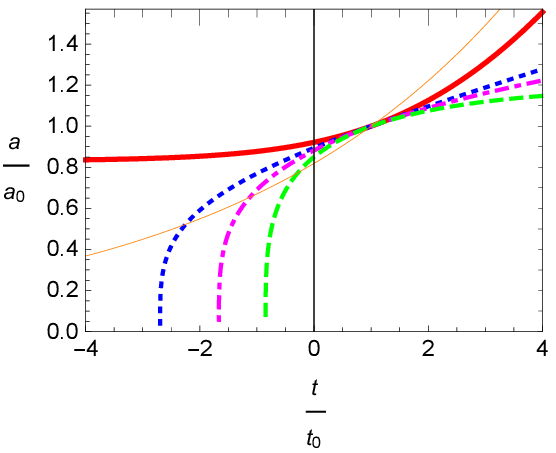}
		% 	\caption{Evolution of $f(T)$: $\mu_1 = 0.45$, $H_0 = 1$, $a_0=\frac{1}{6}$, $H_1=0.5$, $t_0=\frac{1}{6}$, $t_2=2.5$, $x_3=0$ and $x_4=-5.75$}
	\end{minipage}
	% \end{figure}
	% \begin{figure}
	\begin{minipage}{0.495\textwidth}
		\includegraphics[height=.7\textwidth,width=.95\textwidth]{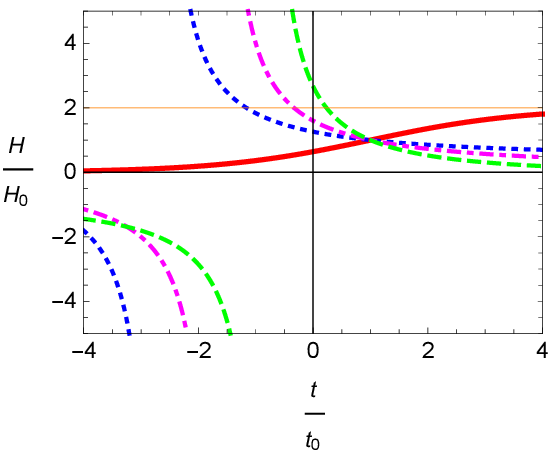}
	\end{minipage}
\begin{minipage}{0.9\textwidth}
\caption{Evolution of $a(t)$ (left panel) and $H(t)$ (right panel) with cosmic time t in early Universe: $H_0 = 1$, $a_0=.1$, $t_0=.1$. Solid line, dotted line, thin line, dot-dashed line and dotted line represent the solutions for equation (\ref{eq19}), equation (\ref{eq17}) with $\beta=\frac{3\gamma H_0}{2}$ and $\beta=3\gamma H_0$,  equation (\ref{eq20}) and equation (\ref{eq21}) with $\beta=-3\gamma H_0$ respectively.}
\label{g1}
\end{minipage}		
\end{figure}

Also $\beta=0$ (i.e. $b=0$) and $\beta(=-\nu^2)<0$ (i.e. $\mu<0$) represent big bang solutions and the solutions are as follows:
\begin{eqnarray}\label{eq20}
\frac{H_0}{H}&=&1+3\gamma H_0 (t-t_0), \nonumber\\
\frac{a}{a_0}&=&\left[1+3\gamma H_0 (t-t_0)\right]^{\frac{1}{3\gamma}}.
\end{eqnarray}

and\vspace{-.5cm}
\begin{eqnarray}\label{eq21}
\frac{H}{H_0}&=&\frac{\nu^2}{\left(\nu^2+3\gamma H_0\right)e^{\nu^2 (t-t_0)}-3\gamma H_0} ,\nonumber\\
\frac{a}{a_0}&=&\left[1+\frac{3\gamma H_0}{\nu^2} \left(1-e^{-\nu^2(t-t_0)}\right)\right]^{\frac{1}{3\gamma}},
\end{eqnarray}

where $t_0$, $a_0$, $H_0$ and $\rho_0$ are integration constants, $a=a_0$, $H=H_0$, $\rho=\rho_0\left(=\sqrt{6}d H_0\right)$ at $t=t_0$. Hence it is possible to have an emergent scenario of the Universe in $f(T)$ gravity theory. Also the cosmic evolution described above (in equations (\ref{eq17})-(\ref{eq21})) are graphically presented in FIG \ref{g1}.
\section{$f(T)$ gravity and Einstein gravity: A correspondence}
% From the usual Friedmann equations with particle creation mechanism the cosmic evolution can be described as \cite{Chakraborty:2014oya}
%\begin{equation}\label{eq22}
%\frac{2\dot H}{3H^2}=\gamma\left(\frac{\Gamma}{3H}-1\right),
%\end{equation}
%
%where $\Gamma$ is the particle creation rate of the cosmic fluid particles. Comparing equation (\ref{eq14}) with equation (\ref{eq22}), one has
%\begin{equation}\label{eq23}
%\frac{\Gamma}{3H}=\frac{\frac{1}{6H^2}\left(Hf_H-f\right)-\frac{1}{12}f_{HH}}{\left(1-\frac{1}{12}f_{HH}\right)},
%\end{equation}
%
%So $f(T)$ gravity can be considered as the particle creation mechanism in Einstein gravity and the particle creation rate is fully depending on $f(T)$.

The modified Friedmann equations in $f(T)$ gravity given by equations (\ref{eq9}) and (\ref{eq10}) can be written in the following forms
\begin{eqnarray}
3H^2&=&\rho-\frac{f}{2}+Tf_T\nonumber \\&=&(\rho+\rho_e)=\rho_T,\label{eq24}\\
2\dot{H}&=&-\gamma\rho-2\dot{H}(f_T+2Tf_{TT})\nonumber \\&=&-\left[(p+\rho)+(p_e+\rho_e)\right]=- (p_T+\rho_T),\label{eq25}
\end{eqnarray}

where $\rho_e,p_e$ are the energy density and thermodynamic pressure of the effective fluid and are given by,
\begin{eqnarray}
\rho_e&=&Tf_T-\frac{f}{2},\label{eq26}\\
p_e&=&2\dot{H}(f_T+2Tf_{TT})-Tf_T+\frac{f}{2}.\label{eq27}
\end{eqnarray}

From the field equations (\ref{eq24}) and (\ref{eq25}) due to Bianchi identity one has,
\begin{equation}
\dot{\rho_T}+3H(p_T+\rho_T)=0,\label{eq28}
\end{equation}

%with $\omega_t=\frac{p_t}{\rho_t}$, as the total equation of state parameter.

From equations (\ref{eq11}) and (\ref{eq28}) one has the individual matter conservation equations as,
\begin{eqnarray}
\dot{\rho}+3H(p+\rho)&=&0,\label{eq29}\\
\dot{\rho_e}+3H(p_e+\rho_e)&=&0.\label{eq30}
\end{eqnarray}

Thus the modified Friedmann equations in $f(T)$ gravity can be interpreted as Friedmann equations in Einstein gravity for an non-interacting two fluid system of which one is the usual normal fluid under consideration and other is the effective fluid. 

%Now equation (\ref{eq30}), the conservation for effective fluid can be rewritten as
%\begin{equation}
%\dot{\rho_e}+3H(\rho_e+p_{e0}+p_{ec})=0,\label{eq31}
%\end{equation}
%
%where $p_{e0}$, $p_{ec}$ are thermodynamic pressure and dissipative pressure of the effective fluid particle with   $\gamma_e=\frac{p_{e0}}{\rho_e}$,  $\omega_e=\frac{p_e}{\rho_e}$. 
%%The variation of $\omega_e$ and $\omega_t$ are shown in FIG. 2 for different choices of $\gamma$.
%
% Due to this dissipative pressure non-equilibrium thermodynamics comes into pictures and one may assume that this dissipative pressure is caused by particle creation process.
%
% So the particle number conservation (of the effective fluid) equation takes the form,
%\begin{equation}
%\dot{n_e}+3Hn_e=\Gamma_e n_e.\label{eq32}
%\end{equation}
%
%where  $n_e$ represents the number density of effective fluid particles.
%
%If the thermodynamical process is chosen to be adiabatic (isentropic) then the dissipative pressure is related to the particle creation rate linearly as \cite{Pan:2014lua},
%\begin{equation}\label{eq33}
%p_{ce}=\frac{\Gamma_e}{3H}\left(p_e+\rho_e\right).
%\end{equation}
%
%From equation (\ref{eq33}) one has,
%\begin{equation}
%\Gamma_e=3H~\frac{\omega_e-\gamma_e}{1+\gamma_e}.\label{eq34}
%\end{equation}
%
%Thus effective particles are created ($\Gamma_e>0$) or annihilated ($\Gamma_e<0$) according as $\omega_e\gtrless\gamma_e$.
%
%
%Further, because of particle creation mechanism there is an energy transfer between the two fluid systems. So these two systems may have different temperatures.

Further using Euler's thermodynamical equation, the evolution of the temperature of the individual fluid are given by \cite{Saha:2014uwa},
\begin{eqnarray}
\frac{\dot{T}}{T}&=&-3H\omega+\frac{\dot{\omega}}{1+\omega},\label{eq35}\\
\frac{\dot{T_e}}{T_e}&=&-3H\omega_e+\frac{\dot{\omega}_e}{1+\omega_e},\label{eq36}
\end{eqnarray}

Integrating equations (\ref{eq35}) and (\ref{eq36}) we have,
\begin{eqnarray}
T&=&T_0 (1+\omega)\exp\left[-3\int_{a_0}^{a} \omega  \frac{da}{a}\right],\label{eq37}\\
T_e&=&T_{0e} (1+\omega_e)\exp\left[-3\int_{a_0}^{a} \omega_e \frac{da}{a}\right],\label{eq38}
\end{eqnarray}

where $T_0$ and $T_{0e}$ are constants of integration. Here $a_0$ is the value of the scale factor at the equilibrium era with equilibrium temperature
$$T_E=T_0(1+\omega_0)=T_{0e}(1+\omega_{0e})$$

where $\omega_0$ and $\omega_{0e}$ are respectively the values of the equation of state parameters $\omega$ and $\omega_e$ at the equilibrium epoch. In particular, using equations (\ref{eq37}), the temperature of the normal fluid for constant $\omega$ can be written explicitly in the following form,
\begin{eqnarray}\label{eq39}
T&=&T_0 (1+\omega)  \left(\frac{a}{a_0}\right)^{-3\omega} .
\end{eqnarray}

\begin{figure}[h]
%	\begin{minipage}{0.49\textwidth}
		\includegraphics[
		height=.35\textwidth,width=.45\textwidth
		]{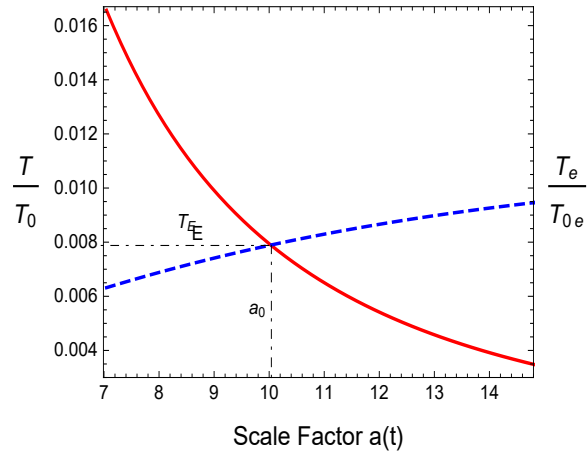}
		% 	\caption{Evolution of $f(T)$: $\mu_1 = 0.45$, $H_0 = 1$, $a_0=\frac{1}{6}$, $H_1=0.5$, $t_0=\frac{1}{6}$, $t_2=2.5$, $x_3=0$ and $x_4=-5.75$}
%	\end{minipage}
	% \end{figure}
	% \begin{figure}
%	\begin{minipage}{0.49\textwidth}
%		\includegraphics[height=.3\textheight,width=.33\textheight]{H.pdf}
%	\end{minipage}
\begin{minipage}{.6\textwidth}
\caption{The figure shows the variation of the temperature $T$ and $T_e$ with the evolution and the equilibrium temperature is identified}	\label{g2}
\end{minipage}	

\end{figure}

In general, at very early phases of the evolution of the universe  $T_e<T$ and then with the evolution of the Universe $T$ decreases while $T_e$ increases and then a state of thermal equilibrium will occur at $T=T_e=T_E$ at $a=a_0$. In the next phase of evolution of universe one has $a>a_0$  and $T_e>T$ because energy flows from effective fluid to the usual fluid continuously and hence the thermodynamical equilibrium is violated. FIG (\ref{g2}) shows the variation of the temperature $T$ and $T_e$ with the scale factor.
 Now, from thermodynamical consideration, equilibrium temperature $T_E$ can be considered as the (modified) Hawking temperature \cite{Chakraborty:2012cw} i.e,
\begin{equation}\label{eq40}
T_E=\frac{H^2 R_h}{2\pi}\bigg|_{a=a_0},
\end{equation}

where $R_h$ is the geometric radius of the horizon, bounding the universe.

Therefore, the present modified gravity theory (i.e. $f(T)$ gravity) is equivalent to Einstein gravity with non-interacting two fluid system having different temperature and the above (modified) Hawking temperature gives the common temperature in equilibrium prescription.
%\begin{figure}
%	\begin{minipage}{0.49\textwidth}
%		\includegraphics[height=.3\textheight,width=.35\textheight]{omegae.eps}
%		% 	\caption{Evolution of $f(T)$: $\mu_1 = 0.45$, $H_0 = 1$, $a_0=\frac{1}{6}$, $H_1=0.5$, $t_0=\frac{1}{6}$, $t_2=2.5$, $x_3=0$ and $x_4=-5.75$}
%	\end{minipage}
%	% \end{figure}
%	% \begin{figure}
%	\begin{minipage}{0.49\textwidth}
%		\includegraphics[height=.3\textheight,width=.35\textheight]{omegat.eps}
%	\end{minipage}
%	\label{f2}
%	\caption{Evolution of $\omega_e$ (left panel) and $\omega_t$ (right panel) with cosmic time t for three different choices of $\gamma$ and for the choices  $\mu_1 = 0.45$, $H_0 = 8$, $H_1=5$, $a_0=\frac{1}{6}$, $t_0=\frac{1}{6}$, $t_2=2.5$, $x_3=0$ and $x_4=-5.75$.}
%\end{figure}
\section{Continuous cosmic evolution in $f(T)$ cosmology}
Now after emergent scenario the Universe evolves through three phases, namely, inflationary era, radiation and matter dominated era, and present time accelerating phase. The motivation of this section comes from \cite{Chakraborty:2014oya} where continuous cosmic evolution has been shown in Einstein gravity with particle creation mechanism. Depending on the choices of the particle creation rate (which are motivated from thermodynamical point of view) it has been shown that there is a continuous cosmic evolution from inflation to the present accelerating era. In this section we shall examine whether such type of continuous cosmic evolution is possible in $f(T)$ cosmology. As there is no physical reason of choosing any form of $f(T)$ so in the following we shall choose $f(T)$ phenomenologically for different eras comparing with the solution equations (in different eras) with that equation (\ref{eq14}). Let $t=t_1$ and $t=t_2(>t_1)$ be the time instants when the Universe evolves from inflationary era to matter dominated era and then to late time accelerating phase .

For inflationary era ($t<t_1$):

Choice for $f(T)$ is
$$f(T)=\frac{\alpha_1 \sqrt{|T|}+\alpha_2 T+ \frac{\mu_1}{\sqrt{6}H_1} |T|^\frac{3}{2}}{\frac{\mu_1}{\sqrt{6}H_1}\sqrt{|T|}-1}$$

and the cosmological solutions are given by
\begin{eqnarray}\label{eq41}
H&=&\frac{H_1}{\mu_1+(1-\mu_1)\left(\frac{a}{a_1}\right)^{\frac{3\gamma}{2}}}\nonumber,\\
a&=&a_1\left[\frac{\mu_1}{1-\mu_1}LambertW\left(\frac{1-\mu_1}{\mu_1}\exp\left\{\frac{2(1-\mu_1)+3\gamma H_1 (t-t_1)} {2\mu_1}\right\} \right)\right]^{\frac{2}{3\gamma}},\nonumber\\
\rho&=&\rho_1 \left(\frac{a}{a_1}\right)^{-3\gamma}.\nonumber\\
q&=&\frac{3\gamma}{2}-1-\mu_1\frac{H}{H_1},
\end{eqnarray}

where $LambertW(x) e^{LambertW(x)}=x$ and $\alpha_1$, $\alpha_2$,  are arbitrary constants. 
 
Here $H_1$, $a_1$, $\rho_1\left(=\frac{\left(6-6\alpha_2+\frac{\sqrt{6}\alpha_1 \mu_1}{H_1}\right)H_1^2}{2(\mu_1-1)^2}\right)$ are the values of the Hubble parameter, scale factor, energy density respectively at $t=t_1$, the time instant at which the universe evolves from inflationary era to matter dominated era and $q=-\left(1+\frac{\dot{H}}{H^2}\right)$ is the deceleration parameter. 

For matter dominated era ($t_1<t<t_2$):

Choice for $f(T)$ is
$$f(T)=\beta_1 \sqrt{|T|}+\beta_2 |T|^{\frac{1}{1-\mu_2}}-T$$

and the cosmological solutions are given by
\begin{eqnarray}\label{eq42}
H&=&H_1\left(\frac{a}{a_1}\right)^{-\frac{3\gamma}{2}(1-\mu_2)}\nonumber,\\
a&=&a_1\left[1+\frac{3\gamma}{2}H_1 (1-\mu_2)(t-t_1)\right]^{\frac{2}{3\gamma(1-\mu_2)}},\nonumber\\
\rho&=&\rho_1 \left(\frac{a}{a_1}\right)^{-3\gamma}\nonumber\\.
q&=&\frac{3\gamma}{2}-1-\mu_2.
\end{eqnarray}

where $\rho_1=\frac{\beta_2}{2}\left(\frac{\mu_2+1}{\mu_2-1}\right)\left(6H_1^2\right)^{\frac{1}{1-\mu_2}}$ and $\beta_1$, $\beta_2$ are arbitrary constants.

For present time accelerating era ($t>t_2$)

Choice for $f(T)$ is
$$f(T)=\delta_1 \sqrt{|T|}+\delta_2T-6(\delta_2+1 )\mu_2H_2^2$$

and the cosmological solutions are given by
\begin{eqnarray}\label{eq43}
H&=&H_2\left[\mu_2 +(1-\mu_2) \left(\frac{a}{a_2}\right)^{-3\gamma} \right]^{\frac{1}{2}},\nonumber\\
a&=&a_2\left[\sqrt{\frac{1-\mu_2}{\mu_2}}\sinh\left\{\frac{3\gamma}{2}\sqrt{\mu_2} H_2 (t-t_i)\right\}\right]^{\frac{2}{3\gamma}},\nonumber\\
\rho&=&\rho_2 \left(\frac{a}{a_2}\right)^{-3\gamma}.\nonumber\\
q&=&\frac{3\gamma}{2}-1-\mu_2 \left(\frac{H_2}{H}\right)^2.
\end{eqnarray}

where $H_2$, $a_2$, $\rho_2 \left(=3\left(\delta_2+1\right)(1-\mu_2) H_2^2\right)$ are the values of the Hubble parameter, scale factor, energy density respectively at $t=t_2$, the time instant at which the universe makes transition from matter dominated era to the present time accelerating era and $\delta_1$, $\delta_2$ are arbitrary constants.

Now smoothness of the deceleration parameter demands $\mu_1=\mu_2$. Also the continuity of the Hubble parameter, scale factor at $t=t_1$ is obvious and continuity across $t=t_2$ leads to the following conditions
\begin{eqnarray}\label{eq44}
\frac{3\gamma}{2}(1-\mu_2)(t_2-t_1)=\frac{1}{H_2}-\frac{1}{H_1}~~~~,~~~~\left(\frac{a_1}{a_2}\right)^{\frac{3\gamma(1-\mu_2)} {2}}=\frac{H_2}{H_1},
\end{eqnarray}\vspace{-.8cm}
\begin{equation}
\sinh(\eta_2)=\sqrt{\frac{\mu_2}{1-\mu_2}},
\end{equation}\label{eq45}
\mbox{~~where}\vspace{-.5cm}
\begin{equation}
\eta_2=\frac{3\gamma}{2}\sqrt{\mu_2} H_2 (t_2-t_i).\nonumber
\end{equation}

Further, let $t=t_0(<t_1)$ be the time instant in which the universe makes a transition from emergent scenario to inflationary era. Then from the continuity across the transition time $t=t_0$, one have
\begin{equation}\label{eq46}
\frac{H_1}{H_0}=\mu_1+(1-\mu_1)\left(\frac{a_0}{a_1}\right)^{\frac{3\gamma}{2}},
\end{equation}

Now, the continuity of $f(T)$ and the energy density across all the transition points $t_0,~t_1,~t_2$ provides the following conditions
\begin{eqnarray}
\sqrt{6}dH_0&=&\frac{\left(6-6\alpha_2+\frac{\sqrt{6}\alpha_1\mu_1}{H_1}\right)H_1^2}{2(\mu_1-1)^2}\left(\frac{a_0}{a_1} \right)^{-3\gamma}\label{eq47}\\
\frac{\left(6-6\alpha_2+\frac{\sqrt{6}\alpha_1\mu_1}{H_1} \right)}{(\mu_1-1)}&=&\beta_2\left(\mu_2+1\right)\left(6H_1^{2\mu_2}\right)^{\frac{1}{1-\mu_2}}\label{eq48}\\
\beta_2\left(\frac{\mu_2+1}{\mu_2-1}\right)\left(6H_1^2\right)^{\frac{1}{1-\mu_2}}\left(\frac{a_2}{a_1}\right)^{-3\gamma}&=&6\left(\delta_2+1\right)(1-\mu_2) H_2^2\label{eq49}\\
4d+c+d\ln\left(6H_0^2\right)&=&\frac{\alpha_1+(1-\alpha_2)\sqrt{6}H_0}{\frac{\mu_1}{H_1}H_0-1}\label{eq50}\\
\frac{\alpha_1+(1-\alpha_2 )\sqrt{6} H_1}{\mu_1-1}&=&\beta_1 +\beta_2\left(\sqrt{6}H_1\right)^\frac{1+\mu_2}{1-\mu_2}\label{eq51}\\
\beta_1+\beta_2\left(\sqrt{6}H_2\right)^\frac{1+\mu_2}{1-\mu_2}&=&\delta_1 -\sqrt{6}(\delta_2+1) (\mu_2+1)H_2\label{eq52}
\end{eqnarray}

%Since evolution equation (\ref{eq10}) have a term like $f_{TT}$,\textbf{ so continuity of $f_T$ demands}
%\begin{eqnarray}
%c+d+d\ln\left(6H_0^2\right)&=&\frac{-\alpha_1+(1-\alpha_2)\left(\frac{\mu_1}{H_1}H_0-2\right)\sqrt{6}H_0}{\left(\frac{\mu_1}{H_1}H_0-1\right)^2}\label{eq53}\\
%\frac{-\alpha_1+(1-\alpha_2)\left(\mu_1-2\right)\sqrt{6}H_1}{\left(\mu_1-1\right)^2}&=&\beta_1 +\frac{2\beta_2}{1-\mu_2} \left(\sqrt{6}H_1\right)^{\frac{1+\mu_2}{1-\mu_2}}\label{eq54}\\
%\beta_1 +\frac{2\beta_2}{1-\mu_2} \left(\sqrt{6}H_2\right)^{\frac{1+\mu_2}{1-\mu_2}}&=&\delta_1 -2(\delta_2+1)\sqrt{6}H_2\label{eq55}
%\end{eqnarray}

\begin{figure}[h]
%	\begin{minipage}{0.49\textwidth}
		\includegraphics[height=.3\textheight,width=.5\textheight]{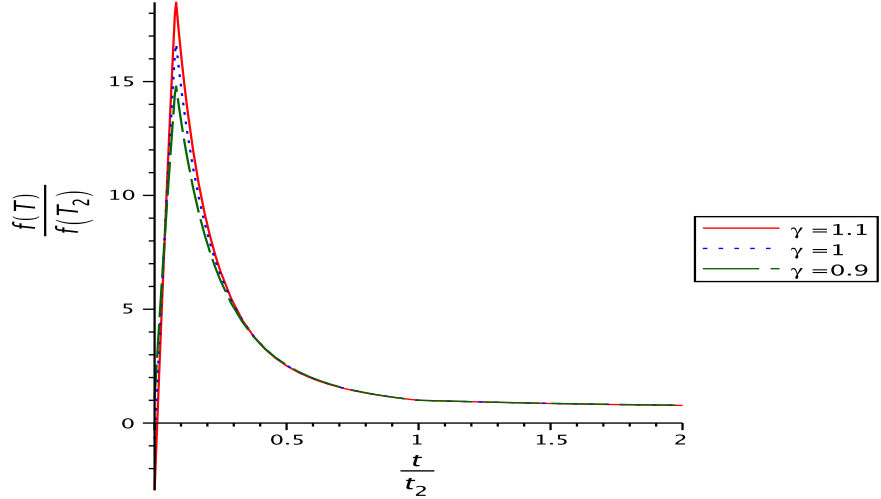}
		% 	\caption{Evolution of $f(T)$: $\mu_1 = 0.45$, $H_0 = 1$, $a_0=\frac{1}{6}$, $H_1=0.5$, $t_0=\frac{1}{6}$, $t_2=2.5$, $x_3=0$ and $x_4=-5.75$}
%	\end{minipage}
	% \end{figure}
	% \begin{figure}
%	\begin{minipage}{0.49\textwidth}
%		\includegraphics[height=.3\textheight,width=.35\textheight]{fplot1.eps}
%	\end{minipage}
\begin{minipage}{0.79\textwidth}
\caption{Evolution of $f(T)$ with cosmic time t for three different choices of $\gamma$ and for the choices: $\mu_1 = 0.25$, $H_0 = 1$, $a_0=1$, $t_0=.2$, $t_2=2.5$, $H_1=0.6$ and $T_2=-6H_2^2$ .}	\label{f3}
\end{minipage}	

\end{figure}

It can be observed that the unknown 8 arbitrary constants: $d,~c,~\alpha_1,~\alpha_2,~\beta_1,~\beta_2,~\delta_1,~\delta_2$ can be determined by using the equations (\ref{eq47})-(\ref{eq52}). So one can easily observe that 2 arbitrary constants must be free. But there are some restrictions on the constants as follows:
$$d>0,~ 6-6\alpha_2+\frac{\sqrt{6}\alpha_1 \mu_1}{H_1}>0, ~\beta_2<0,~ \delta_2>-1.$$
Finally the continuity of $f(T)$ with respect to cosmic time $t$ is also shown graphically in FIG.  \ref{f3} with the choices $\alpha_1=0.3$ and $\alpha_2=-0.5$.
\section{Summary}
A detailed cosmological analysis in $f(T)$ gravity theory has been presented in this work. Firstly, (in section III) it is examined that whether a non-singular model of the universe is possible or not and it is found that for a particular choice of $f(T)$ (given in equation (\ref{eq15})) an emergent scenario is possible only at the level of background cosmology. Note that this choice of $f(T)$ contains terms like square root of $T$ (the $H$ term) and the logarithm of $T$. Hence perturbations must be problematic around emergence of space-time when $T$ tends to zero. Further it is shown that the present $f(T)$ gravity model has a correspondence with the Einstein gravity for non-interacting two fluid system of which one is the usual normal fluid and other is the effective fluid due to the extra geometric term in the $f(T)$ gravity. Also from thermodynamical consideration the temperature of both the fluid particles are evaluated. Lastly, in section V it is possible to have a complete cosmic evolution starting from emergent scenario to present late time accelerating era through the inflationary epoch and matter dominating phase. Hence both physical and geometrical parameters involved in the solutions are found to be continuous (see the figures \ref{g1}-\ref{f3}) over the entire period of evolution. Therefore, one may conclude that $f(T)$ gravity theory may be considered as an alternative to GR in the context of continuous cosmic evolution as well as to resolve the DE problem.

Last but not least since the model is new in the literature and offers various possibilities including the big bang solution for some choices of the parameters which according to the present observational data is the most preferred cosmological model, thus, it naturally demands  further investigations with the available cosmological datasets such as Supernovae Type Ia, Hubble parameter measurements, Baryon Acoustic Oscillations distance measurements etc.   This might be very interesting because the constraints on the model parameters will actually indicate the preference of the cosmological scenario explored in this work. We hope to address in a forthcoming work.
 \section*{Acknowledgement}
  The author A.B. acknowledges UGC-JRF (ID:1207/CSIRNETJUNE2019) and S.C. thanks Science and Engineering Research Board (SERB), India for awarding MATRICS Research Grant support (File No.MTR/2017/000407). The authors are also thankful to Dr. Supriya Pan for his valuable comments and suggestions.


\begin{thebibliography}{100}
\vspace{-.5cm}
	\bibitem{Riess:1998cb} 
	A.~G.~Riess {\it et al.} [Supernova Search Team],
%	``Observational evidence from supernovae for an accelerating universe and a cosmological constant,''
	Astron.\ J.\  {\bf 116}, 1009 (1998).
%	doi:10.1086/300499
%	[astro-ph/9805201].
	
	\bibitem{Perlmutter:1998np}
	S.~Perlmutter \textit{et al.} [Supernova Cosmology Project],
	%``Measurements of $\Omega$ and $\Lambda$ from 42 high redshift supernovae,''
	Astrophys. J. \textbf{517} (1999), 565-586.
%	doi:10.1086/307221
%	[arXiv:astro-ph/9812133 [astro-ph]].
	
	\bibitem{Spergel:2003cb} 
	D.~N.~Spergel {\it et al.} [WMAP Collaboration],
%	``First year Wilkinson Microwave Anisotropy Probe (WMAP) observations: Determination of cosmological parameters,''
	Astrophys.\ J.\ Suppl.\  {\bf 148}, 175 (2003).
%	doi:10.1086/377226
%	[astro-ph/0302209].

	\bibitem{Tegmark:2003ud} 
	M.~Tegmark {\it et al.} [SDSS Collaboration],
%	``Cosmological parameters from SDSS and WMAP,''
	Phys.\ Rev.\ D {\bf 69}, 103501 (2004).
%	doi:10.1103/PhysRevD.69.103501
%	[astro-ph/0310723].
	
	\bibitem{Eisenstein:2005su} 
	D.~J.~Eisenstein {\it et al.} [SDSS Collaboration],
%	``Detection of the Baryon Acoustic Peak in the Large-Scale Correlation Function of SDSS Luminous Red Galaxies,''
	Astrophys.\ J.\  {\bf 633}, 560 (2005).
%	doi:10.1086/466512
%	[astro-ph/0501171].
	
\bibitem{Capozziello:2003tk}
S.~Capozziello, S.~Carloni and A.~Troisi,
%``Quintessence without scalar fields,''
Recent Res. Dev. Astron. Astrophys. \textbf{1} (2003), 625.
%[arXiv:astro-ph/0303041 [astro-ph]].
	
	\bibitem{Carroll:2003wy}
	S.~M.~Carroll, V.~Duvvuri, M.~Trodden and M.~S.~Turner,
	%``Is cosmic speed - up due to new gravitational physics?,''
	Phys. Rev. D \textbf{70} (2004), 043528.
%	doi:10.1103/PhysRevD.70.043528
%	[arXiv:astro-ph/0306438 [astro-ph]].

\bibitem{Einstein:1928}
A. Einstein, Sitzungsber. \ Preuss. \ Akad. \ Wiss. \ Phys. \ Math. \ KI. \textbf{217} (1928);  \textbf{401}
(1930).	

\bibitem{Einstein:1930}
A. Einstein, Math. \ Annal. \textbf{102} (1930) 685.

\bibitem{Hayashi:1979qx} 
K.~Hayashi and T.~Shirafuji,
%``New General Relativity,''
Phys.\ Rev.\ D {\bf 19}, 3524 (1979)
Addendum: [Phys.\ Rev.\ D {\bf 24}, 3312 (1982)].
%doi:10.1103/PhysRevD.19.3524.

\bibitem{Arcos:2005ec} 
H.~I.~Arcos and J.~G.~Pereira,
%``Torsion gravity: A Reappraisal,''
Int.\ J.\ Mod.\ Phys.\ D {\bf 13}, 2193 (2004).
%doi:10.1142/S0218271804006462
%[gr-qc/0501017].

		\bibitem{Ferraro:2006jd} 
	R.~Ferraro and F.~Fiorini,
%	``Modified teleparallel gravity: Inflation without inflaton,''
	Phys.\ Rev.\ D {\bf 75}, 084031 (2007),
%	doi:10.1103/PhysRevD.75.084031
%	[gr-qc/0610067]. %``On Born-Infeld Gravity in Weitzenbock spacetime,''
%	Phys.\ Rev.\ D
	 {\bf 78}, 124019 (2008).
%	doi:10.1103/PhysRevD.78.124019
%	[arXiv:0812.1981 [gr-qc]].
	
	\bibitem{Fiorini:2009ux} 
	F.~Fiorini and R.~Ferraro,
	%``A Type of Born-Infeld regular gravity and its cosmological consequences,''
	Int.\ J.\ Mod.\ Phys.\ A {\bf 24}, 1686 (2009).
%	doi:10.1142/S0217751X09045236
%	[arXiv:0904.1767 [gr-qc]].
	
	\bibitem{Linder:2010py} 
	E.~V.~Linder,
	%	``Einstein's Other Gravity and the Acceleration of the Universe,''
	Phys.\ Rev.\ D {\bf 81}, 127301 (2010)
%	doi:10.1103/PhysRevD.81.127301,
	Erratum: [Phys.\ Rev.\ D {\bf 82}, 109902 (2010)].
%	doi: 10.1103/PhysRevD.82.109902
%	[arXiv:1005.3039 [astro-ph.CO]].
	
	\bibitem{Finch:2018gkh}
	A.~Finch and J.~L.~Said,
	%``Galactic Rotation Dynamics in f(T) gravity,''
	Eur. Phys. J. C \textbf{78} (2018) no.7, 560.
%	doi:10.1140/epjc/s10052-018-6028-1
%	[arXiv:1806.09677 [astro-ph.GA]].
	
	\bibitem{Capozziello:2018hly}
	S.~Capozziello, O.~Luongo, R.~Pincak and A.~Ravanpak,
	%``Cosmic acceleration in non-flat $f(T)$ cosmology,''
	Gen. Rel. Grav. \textbf{50} (2018) no.5, 53.
%	doi:10.1007/s10714-018-2374-4
%	[arXiv:1804.03649 [gr-qc]].
	
	\bibitem{Basilakos:2013rua}
	S.~Basilakos, S.~Capozziello, M.~De Laurentis, A.~Paliathanasis and M.~Tsamparlis,
	%``Noether symmetries and analytical solutions in f(T)-cosmology: A complete study,''
	Phys. Rev. D \textbf{88} (2013), 103526.
%	doi:10.1103/PhysRevD.88.103526
%	[arXiv:1311.2173 [gr-qc]].
	
	\bibitem{Bamba:2013jqa}
	K.~Bamba, S.~D.~Odintsov and D.~Sáez-Gómez,
	%``Conformal symmetry and accelerating cosmology in teleparallel gravity,''
	Phys. Rev. D \textbf{88} (2013), 084042.
%	doi:10.1103/PhysRevD.88.084042
%	[arXiv:1308.5789 [gr-qc]].
	
	\bibitem{Li:2018ixg}
	C.~Li, Y.~Cai, Y.~Cai and E.~N.~Saridakis,
	%``The effective field theory approach of teleparallel gravity, $f(T)$ gravity and beyond,''
	JCAP \textbf{10} (2018), 001.
%	doi:10.1088/1475-7516/2018/10/001
%	[arXiv:1803.09818 [gr-qc]].
	
	\bibitem{Basilakos:2018arq}
	S.~Basilakos, S.~Nesseris, F.~Anagnostopoulos and E.~Saridakis,
	%``Updated constraints on $f(T)$ models using direct and indirect measurements of the Hubble parameter,''
	JCAP \textbf{08} (2018), 008.
%	doi:10.1088/1475-7516/2018/08/008
%	[arXiv:1803.09278 [astro-ph.CO]].
	
	\bibitem{Harko:2014sja}
	T.~Harko, F.~S.~N.~Lobo, G.~Otalora and E.~N.~Saridakis,
	%``Nonminimal torsion-matter coupling extension of f(T) gravity,''
	Phys. Rev. D \textbf{89} (2014), 124036.
%	doi:10.1103/PhysRevD.89.124036
%	[arXiv:1404.6212 [gr-qc]].
	
	\bibitem{Bengochea:2008gz} 
	G.~R.~Bengochea and R.~Ferraro,
%	``Dark torsion as the cosmic speed-up,''
	Phys.\ Rev.\ D {\bf 79}, 124019 (2009).
%	doi:10.1103/PhysRevD.79.124019
%	[arXiv:0812.1205 [astro-ph]].
	
	\bibitem{Li:2010cg} 
	B.~Li, T.~P.~Sotiriou and J.~D.~Barrow,
%	``$f(T)$ gravity and local Lorentz invariance,''
	Phys.\ Rev.\ D {\bf 83}, 064035 (2011).
%	doi:10.1103/PhysRevD.83.064035
%	[arXiv:1010.1041 [gr-qc]].
	
	\bibitem{Cai:2015emx} 
Y.~F.~Cai, S.~Capozziello, M.~De Laurentis and E.~N.~Saridakis,
%``f(T) teleparallel gravity and cosmology,''
Rept.\ Prog.\ Phys.\  {\bf 79}, no. 10, 106901 (2016).
%doi:10.1088/0034-4885/79/10/106901
%[arXiv:1511.07586 [gr-qc]].

		\bibitem{Darabi:2019qpz} 
	F.~Darabi and K.~Atazadeh,
	%``f(T) quantum cosmology,''
	Phys.\ Rev.\ D {\bf 100}, no. 2, 023546 (2019).
%	doi:10.1103/PhysRevD.100.023546
%	[arXiv:1903.03409 [gr-qc]].
	
		\bibitem{Golovnev:2018wbh} 
	A.~Golovnev and T.~Koivisto,
	%``Cosmological perturbations in modified teleparallel gravity models,''
	JCAP {\bf 1811}, 012 (2018).
%	doi:10.1088/1475-7516/2018/11/012
%	[arXiv:1808.05565 [gr-qc]].
	
	\bibitem{Aviles:2013nga} 
	A.~Aviles, A.~Bravetti, S.~Capozziello and O.~Luongo,
	%``Cosmographic reconstruction of $f(\mathcal{T})$ cosmology,''
	Phys.\ Rev.\ D {\bf 87}, no. 6, 064025 (2013).
%	doi:10.1103/PhysRevD.87.064025
%	[arXiv:1302.4871 [gr-qc]].
	
	\bibitem{Bamba:2012} 
	K.~Bamba, M.~ Jamil, D.~Momeni and R.~Myrzakulov,
	%``Generalized second law of thermodynamics in f(T) gravity with entropy corrections,''
	Astrophys.\ Space\ Sci. {\bf 344}, 259-267 (2013).
%	doi:10.1007/s10509-012-1312-2
%	[arXiv:1202.6114 [physics.gen-ph]].

\bibitem{Mishra:2019vnv} 
S.~Mishra and S.~Chakraborty,
%``Stability and bifurcation analysis of interacting f(T) cosmology,''
Eur.\ Phys.\ J.\ C {\bf 79}, no. 4, 328 (2019).
%doi:10.1140/epjc/s10052-019-6839-8
	
	\bibitem{Biswas:2015cva}
	S.~K.~Biswas and S.~Chakraborty,
	%``Interacting Dark Energy in $f(T)$ cosmology : A Dynamical System analysis,''
	Int. J. Mod. Phys. D \textbf{24} (2015) no.07, 1550046.
%	doi:10.1142/S0218271815500467
%	[arXiv:1504.02431 [gr-qc]].
	
	\bibitem{Chakraborty:2014oya} 
	S.~Chakraborty and S.~Saha,
%	``A complete cosmic scenario from inflation to late time acceleration: Non-equilibrium thermodynamics in the context of particle creation,''
	Phys.\ Rev.\ D {\bf 90}, no. 12, 123505 (2014).
%	doi:10.1103/PhysRevD.90.123505
%	[arXiv:1404.6444 [gr-qc]].
	
	\bibitem{Das:2018dzp} 
	D.~Das, S.~Dutta and S.~Chakraborty,
%	``Cosmological consequences in the framework of generalized Rastall theory of gravity,''
	Eur.\ Phys.\ J.\ C {\bf 78}, no. 10, 810 (2018).
%	doi:10.1140/epjc/s10052-018-6293-z
%	[arXiv:1810.11260 [gr-qc]].
	
	\bibitem{Das:2018bxq} 
	D.~Das, S.~Dutta and S.~Chakraborty,
%	``Cosmic scenarios in $f(R)$ gravity: A complete evolution,''
	Annals Phys.\  {\bf 397}, 410 (2018).
%	doi:10.1016/j.aop.2018.08.020
%	[arXiv:1811.00388 [gr-qc]].
	
	\bibitem{Das:2018bxc} 
	D.~Das, S.~Dutta, A.~Al Mamon and S.~Chakraborty,
%	``Does fractal universe describe a complete cosmic scenario ?,''
	Eur.\ Phys.\ J.\ C {\bf 78}, no. 10, 849 (2018).
%	doi:10.1140/epjc/s10052-018-6322-y
%	[arXiv:1811.09674 [physics.gen-ph]].
	
		\bibitem{Rodrigues:2013iua} 
	M.~E.~Rodrigues, I.~G.~Salako, M.~J.~S.~Houndjo and J.~Tossa,
	%``Locally Rotationally Symmetric Bianchi Type-I cosmological model in $f(T)$ gravity: from early to Dark Energy dominated universe,''
	Int.\ J.\ Mod.\ Phys.\ D {\bf 23}, 1450004 (2014).
%	doi:10.1142/S0218271814500047
%	[arXiv:1308.2962 [gr-qc]].
	
		\bibitem{Elizalde:2008yf} 
	E.~Elizalde, S.~Nojiri, S.~D.~Odintsov, D.~Saez-Gomez and V.~Faraoni,
	%``Reconstructing the universe history, from inflation to acceleration, with phantom and canonical scalar fields,''
	Phys.\ Rev.\ D {\bf 77}, 106005 (2008).
%	doi:10.1103/PhysRevD.77.106005
%	[arXiv:0803.1311 [hep-th]].
	
	\bibitem{Amoros:2013nxa} 
	J.~Amorós, J.~de Haro and S.~D.~Odintsov,
	%``Bouncing loop quantum cosmology from $F(T)$ gravity,''
	Phys.\ Rev.\ D {\bf 87}, 104037 (2013).
%	doi:10.1103/PhysRevD.87.104037
%	[arXiv:1305.2344 [gr-qc]].
	
	
	
	\bibitem{Anagnostopoulos:2019miu} 
	F.~K.~Anagnostopoulos, S.~Basilakos and E.~N.~Saridakis,
	%``Bayesian analysis of $f(T)$ gravity using $f\sigma_8$ data,''
	Phys.\ Rev.\ D {\bf 100}, no. 8, 083517 (2019).
%	doi:10.1103/PhysRevD.100.083517
%	[arXiv:1907.07533 [astro-ph.CO]].
	
		\bibitem{ElHanafy:2019zhr} 
	W.~El Hanafy and G.~G.~L.~Nashed,
	%``Phenomenological Reconstruction of $f(T)$ Teleparallel Gravity,''
	Phys.\ Rev.\ D {\bf 100}, no. 8, 083535 (2019).
%	doi:10.1103/PhysRevD.100.083535
%	[arXiv:1910.04160 [gr-qc]].
	
\bibitem{Cai:2019bdh}
Y.~F.~Cai, M.~Khurshudyan and E.~N.~Saridakis,
%``Model-independent reconstruction of $f(T)$ gravity from Gaussian Processes,''
Astrophys. J. \textbf{888} (2020), 62.
%doi:10.3847/1538-4357/ab5a7f
%[arXiv:1907.10813 [astro-ph.CO]].
	
	\bibitem{Mandal:2017xxi} 
	J.~Das Mandal and U.~Debnath,
	%``Dynamical System Analysis of Interacting Hessence Dark Energy in $f(T)$ Gravity,''
	Adv.\ High Energy Phys.\  {\bf 2017}, 2864784 (2017).
%	doi:10.1155/2017/2864784
%	[arXiv:1707.06013 [gr-qc]].
	
	\bibitem{Saha:2016bjs} 
	P.~Saha and U.~Debnath,
	%``Reconstructions of $f(T)$ Gravity from Entropy Corrected Holographic and New Agegraphic Dark Energy Models in Power-law and Logarithmic Versions,''
	Eur.\ Phys.\ J.\ C {\bf 76}, no. 9, 491 (2016).
%	doi:10.1140/epjc/s10052-016-4324-1
%	[arXiv:1608.03272 [gr-qc]].
	
	\bibitem{Farooq:2013ava} 
	M.~U.~Farooq, M.~Jamil, D.~Momeni and R.~Myrzakulov,
	%``Reconstruction of $f(T)$ and $f(R)$ gravity according to $(m, n)$-type holographic dark energy,''
	Can.\ J.\ Phys.\  {\bf 91}, 703 (2013).
%	doi:10.1139/cjp-2012-0431
%	[arXiv:1306.1637 [astro-ph.CO]].
	
	\bibitem{Astashenok:2013kka} 
	A.~V.~Astashenok,
	%``Effective dark energy models and dark energy models with bounce in frames of $F(T)$ gravity,''
	Astrophys.\ Space Sci.\  {\bf 351}, 377 (2014).
%	doi:10.1007/s10509-014-1846-6
%	[arXiv:1308.0581 [gr-qc]].
	

	
	\bibitem{Cardone:2012xq} 
	V.~F.~Cardone, N.~Radicella and S.~Camera,
	%``Accelerating f(T) gravity models constrained by recent cosmological data,''
	Phys.\ Rev.\ D {\bf 85}, 124007 (2012).
%	doi:10.1103/PhysRevD.85.124007
%	[arXiv:1204.5294 [astro-ph.CO]].
	
	\bibitem{Jamil:2012ju} 
	M.~Jamil, D.~Momeni and R.~Myrzakulov,
	%``Resolution of dark matter problem in f(T) gravity,''
	Eur.\ Phys.\ J.\ C {\bf 72}, 2122 (2012).
%	doi:10.1140/epjc/s10052-012-2122-y
%	[arXiv:1209.1298 [gr-qc]].
	
%	\bibitem{Jamil:2012nm} 
%	M.~Jamil, D.~Momeni, R.~Myrzakulov and P.~Rudra,
%	%``Statefinder Analysis of f(T) Cosmology,''
%	J.\ Phys.\ Soc.\ Jap.\  {\bf 81}, 114004 (2012)
%	doi:10.1143/JPSJ.81.114004
%	[arXiv:1211.0018 [physics.gen-ph]].
	
	\bibitem{Daouda:2011yf} 
	M.~Hamani Daouda, M.~E.~Rodrigues and M.~J.~S.~Houndjo,
	%``Reconstruction of f(T) gravity according to holographic dark energy,''
	Eur.\ Phys.\ J.\ C {\bf 72}, 1893 (2012).
%	doi:10.1140/epjc/s10052-012-1893-5
%	[arXiv:1111.6575 [gr-qc]].
	
	\bibitem{Setare:2011ct} 
	M.~R.~Setare and F.~Darabi,
	%``Power-law solutions in $f(T)$ gravity,''
	Gen.\ Rel.\ Grav.\  {\bf 44}, 2521 (2012).
%	doi:10.1007/s10714-012-1408-6
%	[arXiv:1110.3962 [physics.gen-ph]].
	
%	\bibitem{Cai:2011tc} 
%	Y.~F.~Cai, S.~H.~Chen, J.~B.~Dent, S.~Dutta and E.~N.~Saridakis,
%	%``Matter Bounce Cosmology with the f(T) Gravity,''
%	Class.\ Quant.\ Grav.\  {\bf 28}, 215011 (2011)
%	doi:10.1088/0264-9381/28/21/215011
%	[arXiv:1104.4349 [astro-ph.CO]].
	
	\bibitem{Wang:2011xf} 
	T.~Wang,
	%``Static Solutions with Spherical Symmetry in f(T) Theories,''
	Phys.\ Rev.\ D {\bf 84}, 024042 (2011).
%	doi:10.1103/PhysRevD.84.024042
%	[arXiv:1102.4410 [gr-qc]].
%	
%	\bibitem{Dent:2011zz} 
%	J.~B.~Dent, S.~Dutta and E.~N.~Saridakis,
%	%``f(T) gravity mimicking dynamical dark energy. Background and perturbation analysis,''
%	JCAP {\bf 1101}, 009 (2011)
%	doi:10.1088/1475-7516/2011/01/009
%	[arXiv:1010.2215 [astro-ph.CO]].
	
	\bibitem{Zheng:2010am} 
	R.~Zheng and Q.~G.~Huang,
	%``Growth factor in $f(T)$ gravity,''
	JCAP {\bf 1103}, 002 (2011).
%	doi:10.1088/1475-7516/2011/03/002
%	[arXiv:1010.3512 [gr-qc]].
	
	\bibitem{Chen:2010va} 
	S.~H.~Chen, J.~B.~Dent, S.~Dutta and E.~N.~Saridakis,
	%``Cosmological perturbations in f(T) gravity,''
	Phys.\ Rev.\ D {\bf 83}, 023508 (2011).
%	doi:10.1103/PhysRevD.83.023508
%	[arXiv:1008.1250 [astro-ph.CO]].

\bibitem{Chervon:2014tra} 
S.~V.~Chervon, S.~D.~Maharaj, A.~Beesham and A.~S.~Kubasov,
%``Emergent universe supported by chiral cosmological fields in 5D Einstein-Gauss-Bonnet gravity,''
Grav.\ Cosmol.\  {\bf 20}, 176 (2014)
%doi:10.1134/S0202289314030074
%[arXiv:1405.7219 [gr-qc]].

\bibitem{Chakraborty:2014ora} 
S.~Chakraborty,
%``Is Emergent Universe a Consequence of Particle Creation Process?,''
Phys.\ Lett.\ B {\bf 732}, 81 (2014)
%doi:10.1016/j.physletb.2014.03.028
%[arXiv:1403.5980 [gr-qc]].

	\bibitem{Pan:2014lua} 
	S.~Pan and S.~Chakraborty,
	%``Will There Be Future Deceleration? A Study of Particle Creation Mechanism in Nonequilibrium Thermodynamics,''
	Adv.\ High Energy Phys.\  {\bf 2015}, 654025 (2015).
%	doi:10.1155/2015/654025
%	[arXiv:1404.3273 [gr-qc]].
	
	\bibitem{Saha:2014uwa} 
	S.~Saha, A.~Biswas and S.~Chakraborty,
	%``Particle creation and non-equilibrium thermodynamical prescription of dark fluids for universe bounded by an event horizon,''
	Astrophys.\ Space Sci.\  {\bf 356}, no. 1, 141 (2015).
%	doi:10.1007/s10509-014-2189-z
%	[arXiv:1507.08224 [physics.gen-ph]].
	
	\bibitem{Chakraborty:2012cw} 
	S.~Chakraborty,
%	``Is thermodynamics of the universe bounded by the event horizon a Bekenstein system?,''
	Phys.\ Lett.\ B {\bf 718}, 276 (2012).
%	doi:10.1016/j.physletb.2012.11.021.
%	[arXiv:1206.1420 [gr-qc]].
	
%	\bibitem{Akbar:2006er} 
%	M.~Akbar and R.~G.~Cai,
%	%``Friedmann equations of FRW universe in scalar-tensor gravity, f(R) gravity and first law of thermodynamics,''
%	Phys.\ Lett.\ B {\bf 635}, 7 (2006)
%	doi:10.1016/j.physletb.2006.02.035
%	[hep-th/0602156].

\end{thebibliography}
\end{document}